**An integral-transform approach to the bioheat transfer problems in magnetic hyperthermia**

Kenya Murase

*Department of Medical Physics and Engineering, Division of Medical Technology and Science, Faculty of Health Science, Graduate School of Medicine, Osaka University*
*1-7 Yamadaoka, Suita, Osaka 565-0871, Japan*

Short title: An integral-transform approach to magnetic hyperthermia

Address correspondence to:

Kenya Murase, Dr. Med. Sci., Dr. Eng.
Department of Medical Physics and Engineering, Division of Medical Technology and Science, Faculty of Health Science, Graduate School of Medicine, Osaka University
1-7 Yamadaoka, Suita, Osaka 565-0871, Japan
Tel & Fax: (81)-6-6879-2571,
E-mail: murase@sahs.med.osaka-u.ac.jp

**Abstract**

Our purpose in this study was to present an integral-transform approach to the analytical solutions of the Pennes' bioheat transfer equation and to apply it to the calculation of temperature distribution in tissues in hyperthermia with magnetic nanoparticles (magnetic hyperthermia).

The validity of our method was investigated by comparison with the analytical solutions obtained by the Green's function method for point and shell heat sources and the numerical solutions obtained by the finite-difference method for Gaussian-distributed and step-function sources.

There was good agreement between the radial profiles of temperature calculated by our method and those obtained by the Green's function method. There was also good agreement between our method and the finite-difference method except for the central temperature for a step-function source that had approximately a 0.3% difference. We also found that the equations describing the steady-state solutions for point and shell sources obtained by our method agreed with those obtained by the Green’s function method. These results appear to indicate the validity of our method.

In conclusion, we presented an integral-transform approach to the bioheat transfer problems in magnetic hyperthermia, and this study demonstrated the validity of our method. The analytical solutions presented in this study will be useful for gaining some insight into the heat diffusion process during magnetic hyperthermia, for testing numerical codes and/or more complicated approaches, and for performing sensitivity analysis and optimization of the parameters that affect the thermal diffusion process in magnetic hyperthermia.

## 1 Introduction

Hyperthermia is one of the promising approaches to cancer therapy. The most commonly used heating method in the clinical setting is capacitive heating by use of a radiofrequency (RF) electric field [1]. However, a major technical problem with hyperthermia is the difficulty of heating the targeted tumor to the desired temperature without damaging the surrounding tissues, as the electromagnetic energy must be directed from an external source and penetrate normal tissue. Other hyperthermia modalities including RF ablation and ultrasound hyperthermia have been reported [2, 3], but the efficacies of these modalities depend on the size and depth of the tumor, and disadvantages include a limited ability to target the tumor and control the exposure.

Hyperthermia using magnetic nanoparticles (MNPs) (magnetic hyperthermia) was developed in the 1950s [4] and is still under development for overcoming the above disadvantages [5, 6]. MNPs generate heat in an alternating magnetic field as a result of hysteresis and relaxational losses, resulting in heating of the tissue in which MNPs accumulate [7]. With the development of precise methods for synthesizing functionalized MNPs [8], MNPs with functionalized surfaces, which have high specificity for a tumor tissue, have been developed as heating elements for magnetic hyperthermia [9]. Furthermore, there is renewed interest in magnetic hyperthermia as a treatment modality for cancer, especially when it is combined with other, more traditional therapeutic approaches such as the co-delivery of anticancer drugs or photodynamic therapy [10]. From these aspects, magnetic hyperthermia has received much recent attention.

The bioheat transfer equation proposed by Pennes [11] is the basis for understanding the kinetics of the tumor and tissue heating. The solution of this equation is important both for treatment planning and for the design of new clinical heating systems [12].

Various investigations have attempted to obtain analytical solutions to the Pennes' bioheat transfer equation. Durkee and Antich [13] solved it in one-dimensional multi-region Cartesian and spherical geometry, based on the method of separation of variables and Green's function method. Vyas and Rustgi [14] obtained an analytical solution by using the Green's function method to describe the temperature distribution due to a laser beam with a Gaussian profile. Andra et al. [15] solved it for a constant heat source embedded in an infinite medium without blood perfusion by using the Laplace transform. Deng and Liu [16] derived analytical solutions to the bioheat transfer problems with generalized spatial or transient heating both on the skin surface and inside biological bodies by using the Green's function method. Bagaria and Johnson [17] modeled diseased and healthy tissues as two finite concentric spherical regions and included the blood perfusion effect in both regions. They obtained analytical solutions to the model by separation of variables. Recently, Giordano et al. [18] derived fundamental solutions of the Pennes' bioheat transfer equation in rectangular, cylindrical, and spherical coordinates.

Although the Green's function method is a convenient way to describe thermal problems [19, 20] and has often been applied to solving the bioheat transfer equation as described above [12, 14], it is not rare for its handling to become complicated. Besides the Green's function method, analytical solutions to the Pennes' bioheat transfer equation can be obtained by use of the integral-transform method [19, 20], which is considered to be easier to implement than the Green's function method. To the best of our knowledge, however, there are few studies that have used this approach.

Our purpose in this study was to present an integral-transform approach to the analytical solutions of the Pennes' bioheat transfer equation for the calculation of the temperature distribution in tissues in magnetic hyperthermia and to investigate its validity by comparison with the Green's function method and the finite-difference method for several heat source models.

## 2 Materials and methods

### 2.1 Pennes' bioheat transfer equation

To estimate the temperature distribution in vivo, we solved the Pennes' bioheat transfer equation [11] given by

$$\rho c_p \frac{\partial T}{\partial t} = \nabla \cdot (\kappa \nabla T) + \rho_b c_{pb} \omega_b (T_a - T) + Q_{met} + P \tag{1}$$

where $T$ is the temperature of tissue, $\kappa$ the thermal conductivity of tissue, $\rho_b$ the density of blood, $c_{pb}$ the specific heat of blood, $\omega_b$ the blood perfusion rate, $T_a$ the temperature of arterial blood, $Q_{met}$ the rate of metabolic heat generation, and $P$ the energy dissipation. $\rho$ and $c_p$ are the density and specific heat of tissue, respectively. In this study, it was assumed that the volume flow of blood per unit volume is constant and uniform throughout tissue, which means that $\omega_b$ is constant. Furthermore, the above thermo-physical properties such as $\kappa$ and $c_p$ and $Q_{met}$ were assumed to be constant. Therefore, Eq. (1) is reduced to

$$\rho c_p \frac{\partial T}{\partial t} = \nabla \cdot (\kappa \nabla T) + \rho_b c_{pb} \omega_b (T_c - T) + P \tag{2}$$

where

$$T_c = T_a + \frac{Q_{met}}{\rho_b c_{pb} \omega_b} \tag{3}$$

$T_c$ is considered to be the temperature of tissue in the steady state prior to heating or the core body temperature maintained by the balance between metabolic heat generation and blood perfusion. It should be noted that, when $Q_{met}/(\rho_b c_{pb} \omega_b) \ll T_a$ , $T_c$ can be assumed to be equal to $T_a$, as is often seen in the literature [12, 18]. When we describe Eq. (1) in spherical coordinates, Eq. (1) becomes

$$\rho c_p \frac{\partial T}{\partial t} = \frac{\kappa}{r^2} \frac{\partial}{\partial r}\left(r^2 \frac{\partial T}{\partial r}\right) + \rho_b c_{pb} \omega_b (T_c - T) + P \tag{4}$$

## 2.2 Integral-transform method

Applying the integral transform (Fourier sine transform) to Eq. (4) yields (see Appendix A)

$$T = T_c + \sqrt{\frac{2}{\pi}} \frac{1}{\kappa r} \int_0^\infty \frac{F(\beta)}{\alpha^2 + \beta^2} \left[1 - e^{-\frac{\alpha^2+\beta^2}{K} t}\right] \sin(\beta r) d\beta \tag{5}$$

where $K = \rho c_p/\kappa$ and $\alpha^2 = \rho_b c_{pb} \omega_b/\kappa$, and $F(\beta)$ is given by Eq. (A7). When $r = 0$, using the formula $\lim_{x \to 0} \sin x/x = 1$, Eq. (5) is reduced to

$$T = T_c + \sqrt{\frac{2}{\pi}} \frac{1}{\kappa} \int_0^\infty \frac{\beta F(\beta)}{\alpha^2 + \beta^2} \left[1 - e^{-\frac{\alpha^2+\beta^2}{K} t}\right] d\beta \tag{6}$$

In the steady state, i.e., when $t = \infty$, Eq. (5) is reduced to

$$T = T_c + \sqrt{\frac{2}{\pi}} \frac{1}{\kappa r} \int_0^\infty \frac{F(\beta)}{\alpha^2 + \beta^2} \sin(\beta r) d\beta \tag{7}$$

When $r = 0$ and $t = \infty$, Eq. (6) is reduced to

$$T = T_c + \sqrt{\frac{2}{\pi}} \frac{1}{\kappa} \int_0^\infty \frac{\beta F(\beta)}{\alpha^2 + \beta^2} d\beta \tag{8}$$

As illustrative examples, we considered four heat sources (point, shell, Gaussian-distributed, and step-function sources).

### 2.2.1 Point source

In this case, $P$ is given by

$$P = \frac{P_0}{4\pi r^2} \delta(r) \tag{9}$$

where $\delta(r)$ is a Dirac's delta function and $P_0$ is the point heating energy. For this source, $F(\beta)$ given by Eq. (A7) becomes

$$F(\beta) = \sqrt{\frac{2}{\pi}}\frac{P_0}{4\pi}\beta \tag{10}$$

Substituting Eq. (10) into Eq. (5) and using the formula $\int_0^\infty x\sin(ax)/(b^2+x^2)\,dx = \pi e^{-ab}/2$, we obtain

$$T = T_c + \frac{P_0}{2\pi^2\kappa r}\left[\frac{\pi}{2}e^{-\alpha r} - \int_0^\infty \frac{\beta}{\alpha^2+\beta^2}e^{-\frac{\alpha^2+\beta^2}{K}t}\sin(\beta r)d\beta\right] \tag{11}$$

In the steady state, i.e., when $t = \infty$, Eq. (11) is reduced to

$$T = T_c + \frac{P_0}{4\pi\kappa r}e^{-\alpha r} \tag{12}$$

It should be noted that Eq. (12) is also obtained from Eq. (7), and that Eqs. (11) and (12) have a singularity at $r = 0$, which is represented by the factor $1/r$ in the equations and reveals the highly localized effect of a point source.

### 2.2.2 Shell source

In this case, $P$ is given by

$$P = \frac{P_0}{4\pi r^2}\delta(r - r_0) \tag{13}$$

and $F(\beta)$ given by Eq. (A7) becomes

$$F(\beta) = \sqrt{\frac{2}{\pi}}\frac{P_0}{4\pi r_0}\sin(\beta r_0) \tag{14}$$

Substituting Eq. (14) into Eq. (5) and using the formulae $\sin(ax)\sin(bx) = -[\cos(a+b) - \cos(a-b)]/2$ and $\int_0^\infty \cos(ax)/(b^2+x^2)\,dx = \pi e^{-|a|b}/(2b)$, we obtain

$$\begin{aligned} T = T_c + \frac{P_0}{2\pi^2\kappa r r_0}\Bigg\{ & \frac{\pi}{4\alpha}\left[e^{-\alpha|r-r_0|} - e^{-\alpha|r+r_0|}\right] \\ & - \int_0^\infty \frac{\sin(\beta r_0)}{\alpha^2+\beta^2}e^{-\frac{\alpha^2+\beta^2}{K}t}\sin(\beta r)d\beta\Bigg\} \end{aligned} \tag{15}$$

When $r = 0$, substituting Eq. (14 ) into Eq. (6) yields

$$T = T_c + \frac{P_0}{2\pi^2\kappa r_0}\left[\frac{\pi}{2}e^{-\alpha r_0} - \int_0^\infty \frac{\beta\sin(\beta r_0)}{\alpha^2+\beta^2}e^{-\frac{\alpha^2+\beta^2}{K}t}d\beta\right] \tag{16}$$

In the steady state, Eqs. (15) and (16) become

$$T = T_c + \frac{P_0}{8\pi\kappa r r_0 \alpha}\left[e^{-\alpha|r-r_0|} - e^{-\alpha|r+r_0|}\right] \tag{17}$$

and

$$T = T_c + \frac{P_0}{4\pi\kappa r_0} e^{-\alpha r_0} \tag{18}$$

respectively. Note that Eqs. (17) and (18) are also obtained from Eqs. (7) and (8), respectively.

### 2.2.3 Gaussian-distributed source

In this case, $P$ is given by

$$P = P_0 e^{-\frac{r^2}{r_0{}^2}} \tag{19}$$

where $P_0$ is the maximum value of the energy dissipation at the center and $r_0$ is a radius that is associated with how far from the center the heating is affecting the tissue. For this source, $F(\beta)$ given by Eq. (A7) becomes

$$F(\beta) = \frac{\sqrt{2}}{4} P_0 r_0{}^3 \beta e^{-\frac{\beta^2 r_0{}^2}{4}} \tag{20}$$

Substituting Eq. (20) into Eq. (5) yields

$$T = T_c + \frac{P_0 r_0{}^3}{2\sqrt{\pi}\kappa r} \int_0^\infty \frac{\beta e^{-\frac{\beta^2 r_0{}^2}{4}}}{\alpha^2 + \beta^2}\left[1 - e^{-\frac{\alpha^2+\beta^2}{K}t}\right] \sin(\beta r) d\beta \tag{21}$$

In the steady state, i.e., when $t = \infty$, Eq. (21) is reduced to

$$T = T_c + \frac{P_0 r_0{}^3}{2\sqrt{\pi}\kappa r} \int_0^\infty \frac{\beta e^{-\frac{\beta^2 r_0{}^2}{4}}}{\alpha^2 + \beta^2} \sin(\beta r) d\beta \tag{22}$$

When $r = 0$, Eqs. (21) and (22) become

$$T = T_c + \frac{P_0 r_0{}^3}{2\sqrt{\pi}\kappa} \int_0^\infty \frac{\beta^2 e^{-\frac{\beta^2 r_0{}^2}{4}}}{\alpha^2 + \beta^2}\left[1 - e^{-\frac{\alpha^2+\beta^2}{K}t}\right] d\beta \tag{23}$$

and

$$T = T_c + \frac{P_0 r_0{}^3}{2\sqrt{\pi}\kappa} \int_0^\infty \frac{\beta^2 e^{-\frac{\beta^2 r_0{}^2}{4}}}{\alpha^2 + \beta^2} d\beta \tag{24}$$

respectively.

### 2.2.4 Step-function source

In this case, $P$ is given by

$$P = \begin{cases} P_0 & \text{for } 0 \leq r \leq r_0 \\ 0 & \text{for } r_0 < r < \infty \end{cases} \tag{25}$$

and $F(\beta)$ given by Eq. (A7) becomes

$$F(\beta) = \sqrt{\frac{2}{\pi}} P_0 \left[ \frac{1}{\beta^2} \sin(\beta r_0) - \frac{r_0}{\beta} \cos(\beta r_0) \right] \tag{26}$$

Substituting Eq. (26) into Eq. (5) yields

$$T = T_c + \frac{2P_0}{\pi \kappa r} \int_0^{\infty} \frac{\sin(\beta r_0) - \beta r_0 \cos(\beta r_0)}{\beta^2 (\alpha^2 + \beta^2)} \left[ 1 - e^{-\frac{\alpha^2 + \beta^2}{K} t} \right] \sin(\beta r) d\beta \tag{27}$$

When $r = 0$, Eq. (27) becomes

$$T = T_c + \frac{2P_0}{\pi \kappa} \int_0^{\infty} \frac{\sin(\beta r_0) - \beta r_0 \cos(\beta r_0)}{\beta (\alpha^2 + \beta^2)} \left[ 1 - e^{-\frac{\alpha^2 + \beta^2}{K} t} \right] d\beta \tag{28}$$

## 2.3 Green’s function method

### 2.3.1 Point source

The Green’s function of the Pennes’ bioheat transfer equation for radial flow in an infinite domain in spherical coordinates has been given by Giordano et al. [18]. When using this function, we obtain the temperature for a point source as (see Appendix B)

$$T = T_c + \frac{a P_0}{8 \pi^{3/2} \kappa} \int_0^{t} \frac{e^{-b(t-\tau) - \frac{r^2}{4a(t-\tau)}}}{[a(t-\tau)]^{3/2}} d\tau \tag{29}$$

where $a = \kappa/(\rho c_p)$ and $b = \rho_b c_{pb} \omega_b/(\rho c_p)$. It should be noted that, when $r = 0$, the integral in Eq. (29) diverges to infinity.

In the steady state, i.e., when $t = \infty$, the integral in Eq. (29) has an analytical solution: $2\sqrt{\pi} e^{-\sqrt{b/a} r}/(ar)$. Thus, the steady-state solution obtained by the Greens' function method for a point source becomes

$$T = T_c + \frac{P_0}{4 \pi \kappa r} e^{-\sqrt{\frac{b}{a}} r} \tag{30}$$

### 2.3.2 Shell source

When using the Green's function given by Giordano et al. [18], we obtain the temperature for a shell source as (see Appendix B)

$$T = T_c + \frac{aP_0}{8\pi\kappa r r_0} \int_0^t \frac{e^{-b(t-\tau)}}{\sqrt{a\pi(t-\tau)}} \left[ e^{-\frac{(r-r_0)^2}{4a(t-\tau)}} - e^{-\frac{(r+r_0)^2}{4a(t-\tau)}} \right] d\tau \tag{31}$$

When $r = 0$, we obtain from Eq. (B8)

$$T = T_c + \frac{aP_0}{8\pi^{3/2}\kappa} \int_0^t \frac{e^{-b(t-\tau)-\frac{{r_0}^2}{4a(t-\tau)}}}{[a(t-\tau)]^{3/2}} d\tau \tag{32}$$

In the steady state, i.e., when $t = \infty$, the integral in Eq. (31) has an analytical solution: $\left[ e^{-\sqrt{b/a}|r-r_0|} - e^{-\sqrt{b/a}(r+r_0)} \right] / \sqrt{ab}$. Thus, the steady-state solution obtained by the Greens' function method for a shell source becomes

$$T = T_c + \frac{P_0}{8\pi\kappa r r_0} \sqrt{\frac{a}{b}} \left[ e^{-\sqrt{\frac{b}{a}}|r-r_0|} - e^{-\sqrt{\frac{b}{a}}(r+r_0)} \right] \tag{33}$$

Similarly, the integral at $t = \infty$ in Eq. (32) has an analytical solution: $2\sqrt{\pi} e^{-\sqrt{b/a} r_0} / (a r_0)$. Thus, the steady-state solution for a shell source at $r = 0$ becomes

$$T = T_c + \frac{P_0}{4\pi\kappa r_0} e^{-\sqrt{\frac{b}{a}} r_0} \tag{34}$$

## 2.4 Finite-difference method

We also solved Eq. (4) by using the finite-difference method (forward-difference scheme) (see Appendix C) for Gaussian-distributed and step-function sources for comparison. When we used the finite-difference method (Appendix C), the outer radius of the domain for analysis was taken as 15 cm, and the spatial and time intervals ($\Delta$r and $\Delta$t) were taken as 0.3 mm and $\rho c_p \Delta \mathrm{r}^2 / (2\kappa)$, respectively.

## 2.5 Energy dissipation of magnetic nanoparticles

Rosensweig [7] developed analytical relationships and computations of the energy dissipation of MNPs subjected to an alternating magnetic field (AMF). From this theory, *P* in Eq. (4) can be given by [6, 7]

$$P = \pi \mu_0 \chi_0 {H_0}^2 f \frac{2\pi f \tau}{1 + (2\pi f \tau)^2} \tag{35}$$

where $\mu_0$ is the permeability of free space, $\chi_0$ the equilibrium susceptibility, and $H_0$ and $f$ the amplitude and frequency of the AMF, respectively. $\tau$ is the effective relaxation time given by

$$\frac{1}{\tau} = \frac{1}{\tau_N} + \frac{1}{\tau_B} \tag{36}$$

where $\tau_N$ and $\tau_B$ are the Neel relaxation and Brownian relaxation time, respectively [6, 7]. $\tau_N$ and $\tau_B$ are given by the following relationships [6, 7]:

$$\tau_N = \tau_0 \frac{\sqrt{\pi} e^{\Gamma}}{2\sqrt{\Gamma}} \text{ and } \tau_B = \frac{3\eta V_H}{k_B T} \tag{37}$$

where $\tau_0$ is the average relaxation time in response to a thermal fluctuation, $\eta$ the viscosity of medium, $k_B$ the Boltzmann constant, $T$ the temperature, and $\Gamma = KV_M/(k_B T)$, with $K$ being the anisotropy constant of MNP. $V_H$ is taken as the hydrodynamic volume of MNP that is larger than the magnetic volume $V_M = \pi D^3/6$ for MNP of diameter $D$. As a model for $V_H$, it is assumed that $V_H = (1 + 2\delta/D)^3 V_M$, where $\delta$ is the thickness of a sorbed surfactant layer. Because the actual equilibrium susceptibility $\chi_0$ is dependent on the magnetic field, $\chi_0$ is assumed to be the chord susceptibility corresponding to the Langevin equation, given by

$$\chi_0 = \chi_i \frac{3}{\xi}\left(\coth\xi - \frac{1}{\xi}\right) \tag{38}$$

where $\chi_i = \mu_0 \phi M_d{}^2 V_M/(3k_B T)$, $\xi = \mu_0 M_d H V_M/(k_B T)$, $H = H_0 \cos(2\pi f t)$, $M_d$ is the domain magnetization of a suspended particle, and $\phi$ is the volume fraction of MNPs.

In this study, we considered magnetite ($Fe_3O_4$) as MNPs. The above parameters for magnetite were taken to be as follows: $M_d = 446$ kA/m, $K = 9$ kJ/m$^3$, $c_p = 670$ J/kg/K, and $\rho = 5180$ kg/m$^3$ [21]. $\phi$ was taken as 0.003, which is close to the typical magnetite dosage of ~10 mg Fe per gram of tumor that has been reported in clinical studies [22].

Figure 1(a) shows the relationship between $P$ and $D$ for magnetite, in which $H_0$ was fixed at 5 mT and $f$ was varied from 100 kHz to 1000 kHz with an interval of 100 kHz, whereas Fig. 1(b) shows the case when $f$ was fixed at 500 kHz and $H_0$ was varied from 1 mT to 10 mT with an interval of 1 mT. It should be noted that the unit of mT can be converted to kA/m by use of the relationship 1 mT = 0.796 kA/m. As shown in Fig. 1, $P$ largely depends on $D$ and its maximum value increases with increasing $f$ and $H_0$.

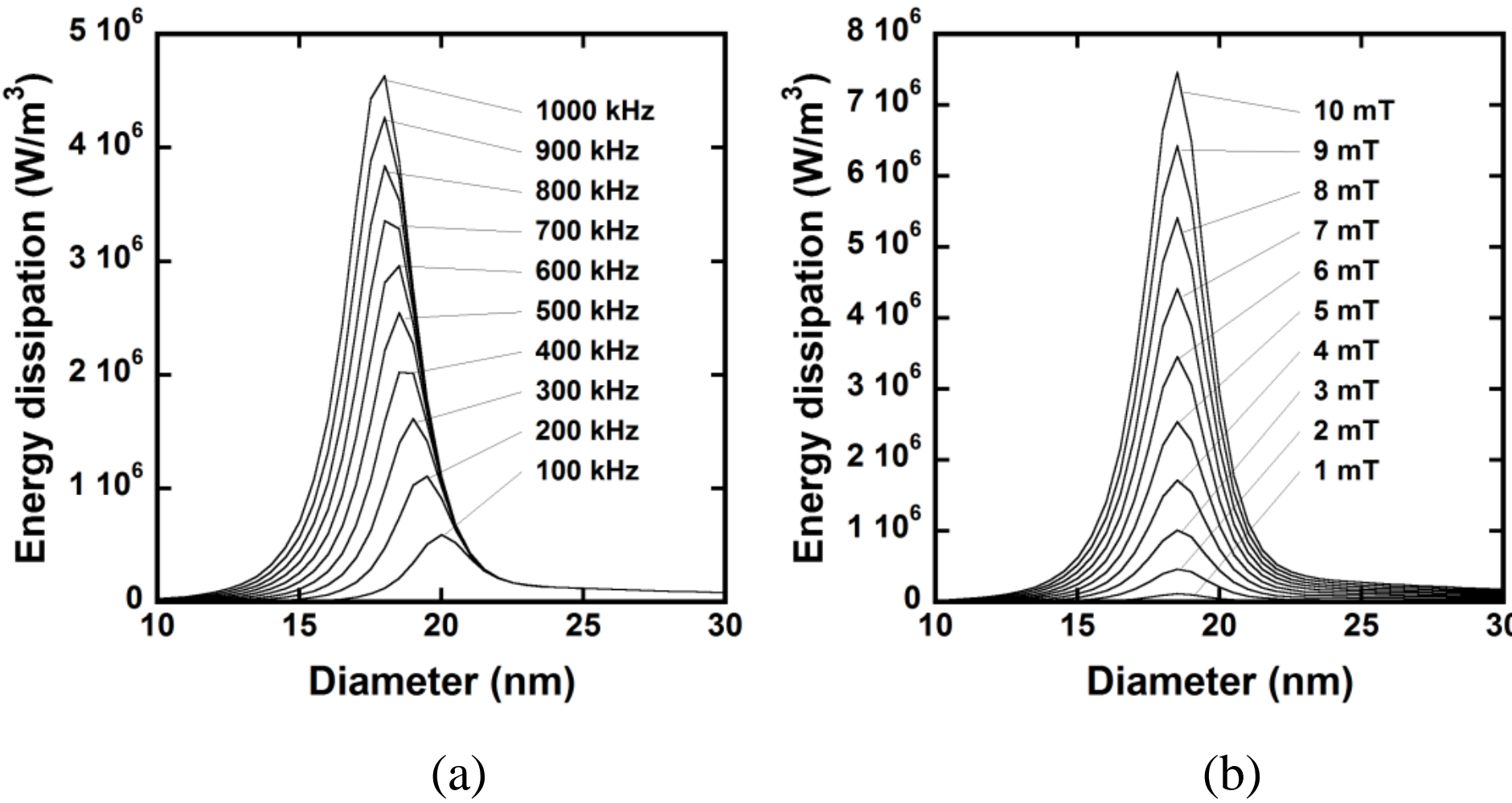

(a) (b)

Fig. 1 (a) Relationship between the energy dissipation ($P$) and the diameter of magnetic nanoparticles (MNPs) ($D$) for magnetite, in which the amplitude of the alternating magnetic field ($H_0$) was fixed at 5 mT and the frequency ($f$) was varied from 100 kHz to 1000 kHz with an interval of 100 kHz. (b) Relationship between $P$ and $D$ for magnetite, in which $f$ was fixed at 500 kHz and $H_0$ was varied from 1 mT to 10 mT with an interval of 1 mT. The unit of mT can be converted to kA/m by use of the relationship $1 \text{ mT} = 0.796 \text{ kA/m}$.

As an illustrative example, we considered the case with $D = 19$ nm, $f = 500$ kHz, and $H_0 = 5$ mT. In this case, $P = 2.28 \times 10^6 \text{ W/m}^3$. For a point source, we assumed that MNPs were located within a sphere with a radius of 1 mm. From the relationship $P = P_0/V$, where $V$ is the volume of the region where MNPs are located, $P_0$ in Eq. (9) was taken as 0.0096 W. For a shell source, $r_0$ and the width of the shell were assumed to be 5 mm and 1 mm, respectively, resulting in $V = 315.2 \text{ mm}^3$. Thus, $P_0$ in Eq. (13) was assumed to be 0.72 W. For Gaussian-distributed and step-function sources, $P_0 = P = 2.28 \times 10^6 \text{ W/m}^3$ was used in Eqs. (19) and (25).

## 2.6 Numerical studies

Numerical studies were performed under the following conditions: The values for the thermo-physical properties of blood and tissue were assumed to be as follows [21]: $\kappa = 0.502 \text{ W/m/K}$, $\rho = 1060 \text{ kg/m}^3$, $c_p = 3600 \text{ J/kg/K}$, $\rho_b = 1000 \text{ kg/m}^3$, $c_{pb} = 4180 \text{ J/kg/K}$, $\omega_b = 6.4 \times 10^{-3} \text{ s}^{-1}$, and $T_c = 310$ K. In this study, the $r_0$ values in Eqs. (13), (19), and (25) were all taken as 5 mm.

## 3 Results

Figure 2 shows a comparison of the radial profiles of temperature calculated by our method and those calculated by the Green's function method for a point source at three time points (5, 10, and 100 s). As shown in Fig. 2, there was good agreement between them.

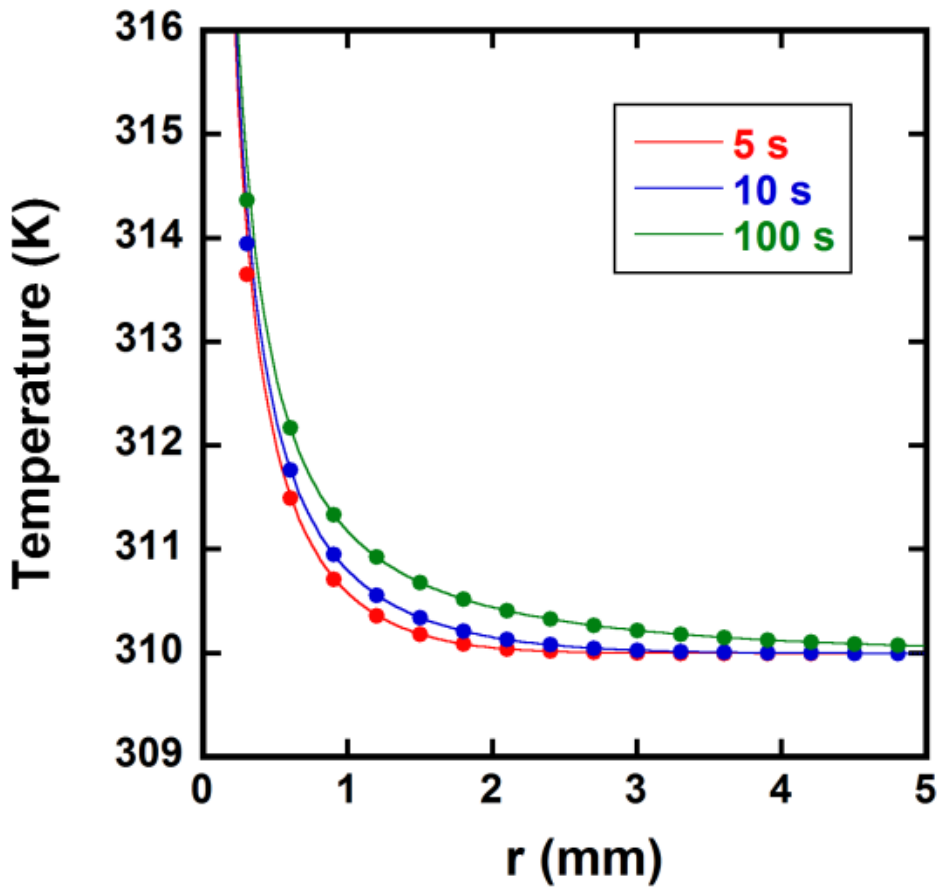

Fig. 2 Comparison of the radial profiles of temperature calculated by our method and those obtained by the Green's function method for a point source. The red, blue, and green solid lines show the results calculated by our method at $t = 5$ s, 10 s, and 100 s, respectively, whereas the red, blue, and green closed circles show the results obtained by the Green's function method at $t = 5$ s, 10 s, and 100 s, respectively.

Figure 3 shows a comparison of the radial profiles of temperature calculated by our method and those calculated by the Green's function method for a shell source at four time points (10, 50, 100, and 1000 s). As shown in Fig. 3, there was good agreement between them.

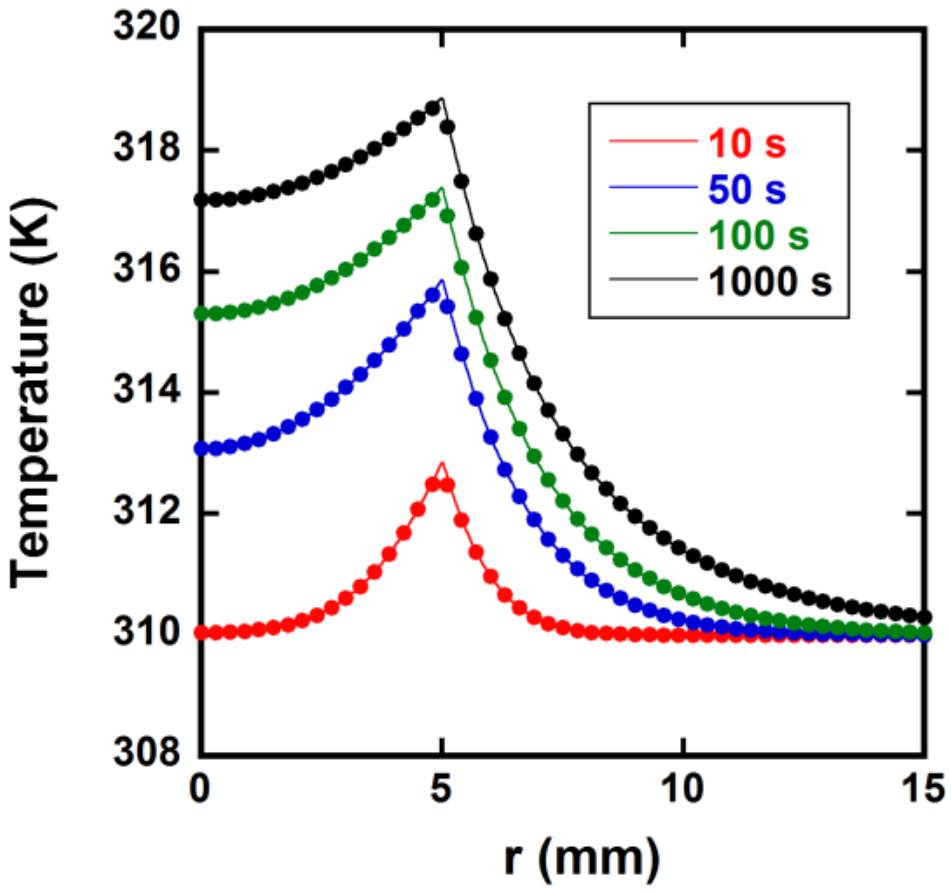

Fig. 3 Comparison of the radial profiles of temperature calculated by our method and those obtained by the Green's function method for a shell source. The red, blue,

green, and black solid lines show the results calculated by our method at $t =$ 10 s, 50 s, 100 s, and 1000 s, respectively, whereas the red, blue, green, and black closed circles show the results obtained by the Green's function method at $t =$ 10 s, 50 s, 100 s, and 1000 s, respectively.

Figure 4 shows a comparison of the radial profiles of temperature calculated by our method and those calculated by the finite-difference method for a Gaussian-distributed source at four time points (10, 50, 100, and 500 s). As shown in Fig. 4, there was good agreement between them.

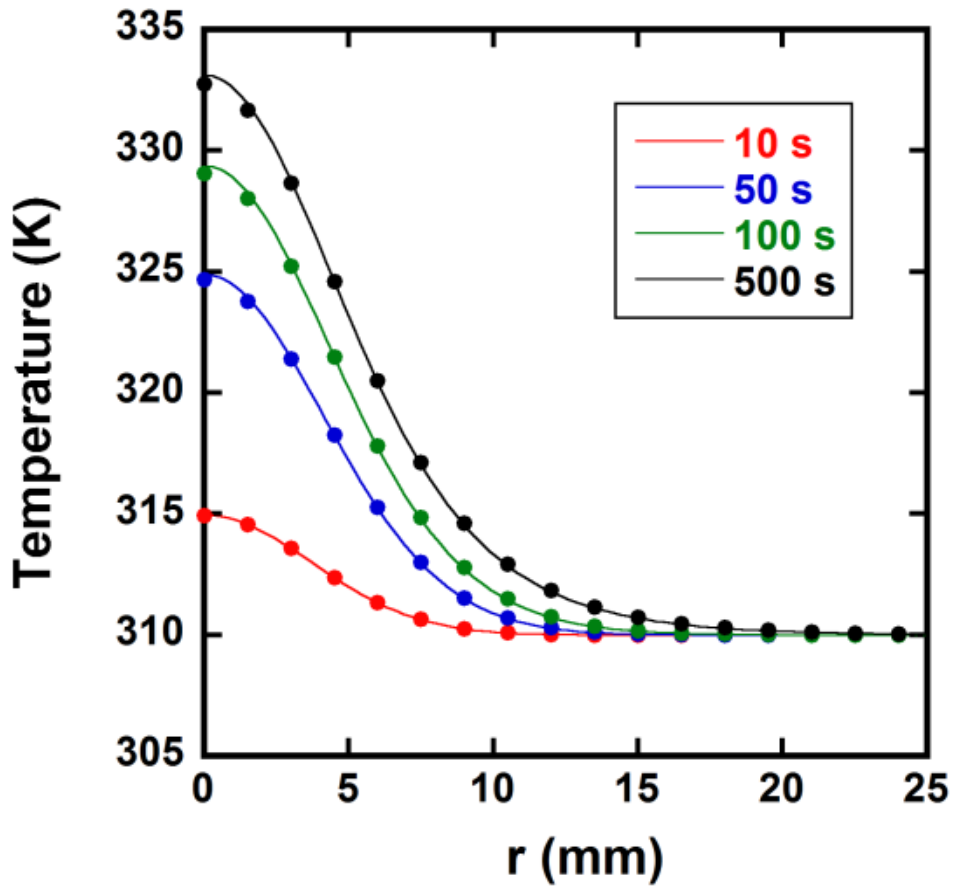

Fig. 4 Comparison of the radial profiles of temperature calculated by our method and those obtained by the finite-difference method for a Gaussian-distributed source. The red, blue, green, and black solid lines show the results calculated by our method at $t =$ 10 s, 50 s, 100 s, and 500 s, respectively, whereas the red, blue, green, and black closed circles show the results obtained by the finite-difference method at $t =$ 10 s, 50 s, 100 s, and 500 s, respectively.

Figure 5 shows a comparison of the radial profiles of temperature calculated by our method and those calculated by the finite-difference method for a step-function source at four time points (10, 50, 100, and 500 s). As shown in Fig. 5, although some difference (approximately 0.3%) was observed at $r = 0$, there was good agreement between them except for the central temperature.

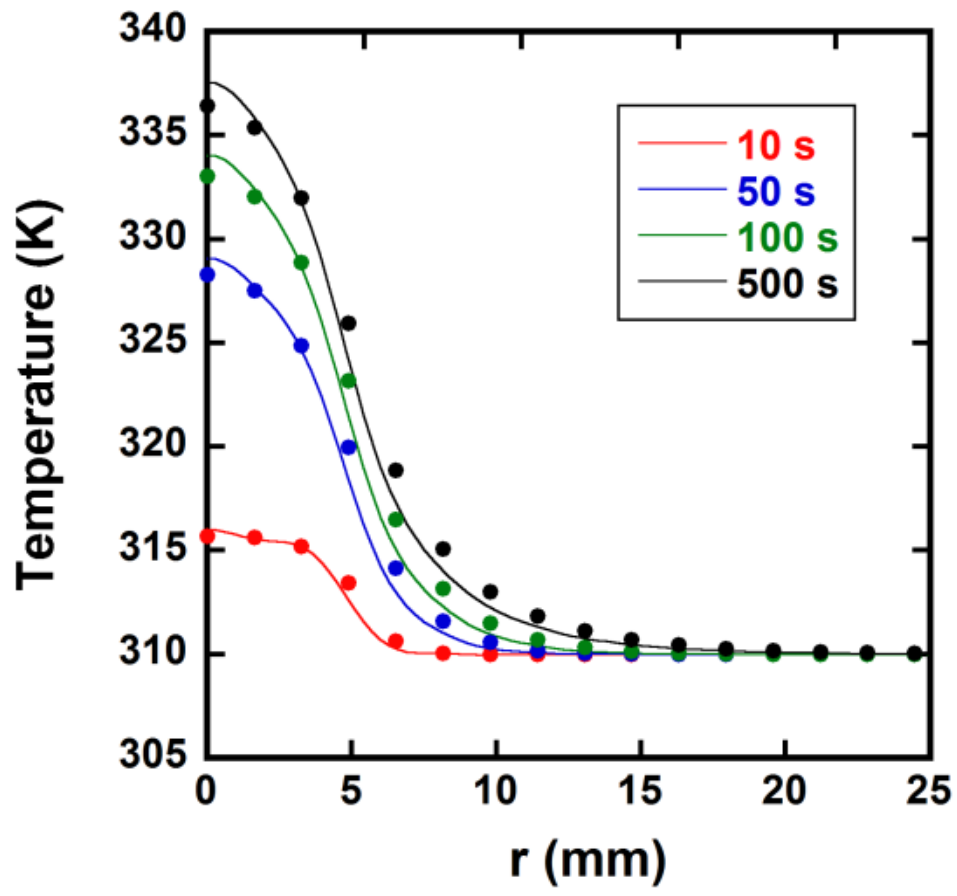

Fig. 5 Comparison of the radial profiles of temperature calculated by our method and those obtained by the finite-difference method for a step-function source. The red, blue, green, and black solid lines show the results calculated by our method at $t =$ 10 s, 50 s, 100 s, and 500 s, respectively, whereas the red, blue, green, and black closed circles show the results obtained by the finite-difference method at $t =$ 10 s, 50 s, 100 s, and 500 s, respectively.

## 4 Discussion

In this study, we presented an integral-transform approach to the bioheat transfer problems in magnetic hyperthermia and derived the transient and steady-state analytical solutions to the Pennes' bioheat transfer equation for several heat source models by using this approach. Furthermore, we investigated the validity of this approach by comparison with the analytical solutions obtained by the Green’s function method for point and shell sources and the numerical solutions obtained by the finite-difference method for Gaussian-distributed and step-function sources. To the best of our knowledge, these analytical solutions obtained by the integral-transform approach have not been reported previously. The largest difference was observed between the central temperature obtained by our method and that obtained by the finite-difference method for a step-function source (Fig. 5), but the difference was approximately 0.3% at most. Excluding this case, there was good agreement between our method and the Green’s function method or the finite-difference method (Figs. 2–5), indicating the validity of our method.

As previously described, the steady-state solutions obtained by the Greens' function method for point and shell sources are given by Eqs. (30) and (33), respectively. Because $\sqrt{b/a} = \alpha$, Eqs. (30) and (33) agree with Eqs. (12) and (17) derived from our method, respectively. Furthermore, the steady-state solution for a shell source at $r = 0$ obtained by the Greens' function method [Eq. (34)] also agrees with that obtained by our method [Eq. (18)]. These results also appear to indicate the validity of our method.

The shell source used in this study is a model consisting of a thin shell of MNPs in the outer surface of a spherical solid tumor whose outer region extends to infinity and represents the normal tissue. As pointed out by Giordano et al. [18], this model is a realistic model distribution that provides an approximately constant therapeutic temperature inside the tumor. For this model, there was also good agreement between our method and the Green's function method.

The Green's function method is a convenient way of solving differential equations such as the Pennes' bioheat transfer equation [12, 14]. Mathematically, the Green's function is the solution to a differential equation with an instantaneous point source. When the temperature distribution for various heat sources is calculated by use of the Green's function method, it is necessary to compute the integral of the product of the Green's function and the function describing the heat source as shown in Eqs. (B4) and (B7). In general, this integral becomes a double integral with respect to the temporal and spatial variables. For point and shell sources that can be described by a Dirac's delta function as shown in Eqs. (9) and (13), it is relatively easy to compute the double integral. However, it is not always easy to compute the double integral for heat sources whose function cannot be described by a Dirac's delta function, such as Gaussian-distributed and step-function sources. On the other hand, the integral-transform method presented in this study appears much easier to implement than the Green's function method.

In the integral-transform method presented in this study, the kernel for the integral transform was taken to be $\sqrt{2/\pi}\sin(\beta r)$. In general, the kernel should be chosen depending on the boundary conditions at $r = 0$ [19]. When the boundary condition at $r = 0$ is of the first kind, the kernel should be $\sqrt{2/\pi}\sin(\beta r)$, whereas it should be $\sqrt{2/\pi}\cos(\beta r)$ for the boundary condition of the second kind [19]. In this study, the parameter $\theta$ [$= \Theta \cdot r$ (see Appendix A)] is always zero at $r = 0$, that is, the boundary condition at $r = 0$ is of the first kind. Thus, we used $\sqrt{2/\pi}\sin(\beta r)$ as the kernel for the integral transform in this study.

The analytical solutions presented in this study were based on several assumptions. First, the domain for analysis was assumed to be infinite. Although this assumption is considered to be valid for deep tumors surrounded by normal tissue, our method cannot be applied to the case of relatively superficial tumors. Second, the thermo-physical properties of blood and tissue were assumed to be the same in both the tumor and normal tissue. Third, the shape of tumors and the distribution of MNPs were assumed to be spherically symmetric. Although the analytical solutions derived in this study cannot be applied to cases with complex geometries and/or a heterogeneous medium, they will provide useful tools for testing of numerical codes and/or more complicated approaches, and for performing sensitivity analysis of the parameters involved in a problem [18].

In conclusion, we presented an integral-transform approach to the bioheat transfer problems in magnetic hyperthermia, and this study demonstrated the validity of our method. The analytical solutions presented in this study will be useful for gaining some insight into the heat-diffusion process during magnetic hyperthermia, for testing of numerical codes and/or more complicated approaches, and for performing sensitivity analysis and optimization of the parameters that affect the thermal diffusion process in magnetic hyperthermia.

**Appendix A**

When the following parameter is introduced:

$$\Theta = T - T_c \tag{A1}$$

Eq. (4) is reduced to

$$\rho c_p \frac{\partial \Theta}{\partial t} = \frac{\kappa}{r^2}\frac{\partial}{\partial r}\left(r^2 \frac{\partial \Theta}{\partial r}\right) - \rho_b c_{pb} \omega_b \Theta + P \tag{A2}$$

Furthermore, if we perform the following variable transformation: $\theta = \Theta \cdot r$, Eq. (A2) becomes

$$K \frac{\partial \theta}{\partial t} = \frac{\partial^2 \theta}{\partial r^2} - \alpha^2 \theta + \frac{rP}{\kappa} \tag{A3}$$

where

$$K = \frac{\rho c_p}{\kappa} \text{ and } \alpha^2 = \frac{\rho_b c_{pb} \omega_b}{\kappa} \tag{A4}$$

If we apply the integral transform (Fourier sine transform) [19, 20] to Eq. (A3), we obtain

$$K \frac{\partial \bar{\theta}}{\partial t} = -(\alpha^2 + \beta^2)\bar{\theta} + \frac{F(\beta)}{\kappa} \tag{A5}$$

where $\bar{\theta}$ is defined by [20]

$$\bar{\theta} = \sqrt{\frac{2}{\pi}} \int_0^\infty \theta \sin(\beta r) \, dr \tag{A6}$$

and

$$F(\beta) = \sqrt{\frac{2}{\pi}} \int_0^\infty r P(r) \sin(\beta r) \, dr \tag{A7}$$

$\beta$ denotes the Fourier-transform variable, which is assumed to take all values from 0 to infinity continuously. It should be noted that $\theta$ and $\partial\theta/\partial r$ at $r = \infty$ were taken as zero to obtain Eq. (A5).

Solving Eq. (A5) with respect to $t$ yields

$$\bar{\theta} = e^{-\frac{\alpha^2+\beta^2}{K}t} \bar{\theta}(0) + \frac{F(\beta)}{\kappa(\alpha^2+\beta^2)} \left[ 1 - e^{-\frac{\alpha^2+\beta^2}{K}t} \right] \tag{A8}$$

where $\bar{\theta}(0)$ is the value of $\bar{\theta}$ at $t = 0$. If we assume that the temperature ($T$) at $t = 0$ is equal to $T_c$, we obtain $\bar{\theta}(0) = 0$. Using the following inverse Fourier transformation [20]:

$$\theta = \sqrt{\frac{2}{\pi}} \int_0^\infty \bar{\theta}(\beta) \sin(\beta r) \, d\beta \tag{A9}$$

we obtain

$$\theta = \sqrt{\frac{2}{\pi}} \frac{1}{\kappa} \int_0^\infty \frac{F(\beta)}{\alpha^2+\beta^2} \left[ 1 - e^{-\frac{\alpha^2+\beta^2}{K}t} \right] \sin(\beta r) \, d\beta \tag{A10}$$

Finally, by use of $\Theta = \theta/r$ and Eqs. (A1) and (A4), the temperature ($T$) can be obtained by Eq. (9).

## Appendix B

The Green's function of Eq. (A2) in an infinite domain is given by [18]

$$G(r, t; r', \tau) = \frac{e^{-b(t-\tau)}}{2rr'\sqrt{a\pi(t-\tau)}} \left[ e^{-\frac{(r-r')^2}{4a(t-\tau)}} - e^{-\frac{(r+r')^2}{4a(t-\tau)}} \right] \tag{B1}$$

where

$$a = \frac{\kappa}{\rho c_p} = \frac{1}{K} \tag{B2}$$

and

$$b = \frac{\rho_b c_{pb} \omega_b}{\rho c_p} = \frac{\alpha^2}{K} \tag{B3}$$

**Point source**

For a point source model, $P$ is given by Eq. (12). In this case, the solution to Eq. (A2) is given by

$$\begin{aligned} \Theta &= \frac{a}{\kappa}\int_0^t \int_0^\infty r'^2\, G(r,t;r',\tau)\frac{P_0\delta(r')}{4\pi r'^2}dr'd\tau \\ &= \frac{aP_0}{8\pi\kappa}\int_0^t \int_0^\infty \frac{e^{-b(t-\tau)}\delta(r')}{rr'\sqrt{a\pi(t-\tau)}}\left[e^{-\frac{(r-r')^2}{4a(t-\tau)}} - e^{-\frac{(r+r')^2}{4a(t-\tau)}}\right]dr'd\tau \end{aligned} \tag{B4}$$

Using the following relationship:

$$e^{-\frac{(r-r')^2}{4a(t-\tau)}} - e^{-\frac{(r+r')^2}{4a(t-\tau)}} = \frac{4\sqrt{a(t-\tau)}}{\sqrt{\pi}}\int_0^\infty e^{-a\lambda^2(t-\tau)}\sin(\lambda r)\sin(\lambda r')d\lambda \tag{B5}$$

and the formula $\int_0^\infty e^{-a^2x^2}\cos(bx)\,dx = \sqrt{\pi}e^{-b^2/(4a^2)}/(2a)$, we obtain

$$\Theta = \frac{aP_0}{8\pi^{3/2}\kappa}\int_0^t \frac{e^{-b(t-\tau)-\frac{r^2}{4a(t-\tau)}}}{[a(t-\tau)]^{3/2}}d\tau \tag{B6}$$

Thus, we obtain Eq. (29).

**Shell source**

In this case, $P$ is given by Eq. (16). In this case, the solution to Eq. (A2) is given by

$$\begin{aligned} \Theta &= \frac{a}{\kappa}\int_0^t \int_0^\infty r'^2\, G(r,t;r',\tau)\frac{P_0\delta(r'-r_0)}{4\pi r'^2}dr'd\tau \\ &= \frac{aP_0}{8\pi\kappa r r_0}\int_0^t \frac{e^{-b(t-\tau)}}{\sqrt{a\pi(t-\tau)}}\left[e^{-\frac{(r-r_0)^2}{4a(t-\tau)}} - e^{-\frac{(r+r_0)^2}{4a(t-\tau)}}\right]d\tau \end{aligned} \tag{B7}$$

Thus, we obtain Eq. (31).

When $r = 0$, using Eq. (B5) and the formula $\lim_{x\to 0}\sin x/x = 1$ yields

$$\Theta = \frac{aP_0}{8\pi^{3/2}\kappa}\int_0^t \frac{e^{-b(t-\tau)-\frac{{r_0}^2}{4a(t-\tau)}}}{[a(t-\tau)]^{3/2}}d\tau \tag{B8}$$

Thus, we obtain Eq. (32).

**Appendix C**

To solve Eq. (4), we used the following finite-difference method (forward-difference scheme). First, we divide the spatial and time domains into small intervals $\Delta r$ and $\Delta t$ such that $r = (i-1)\cdot\Delta r\ (i = 1,2,\cdots,M)$ and $t = (j-1)\cdot\Delta t\ (j = 1,2,\cdots,N)$, and we denote the temperature at the nodal point $i \cdot \Delta r$ at the time $j \cdot \Delta t$ by $T_{i,j}$. For $r \neq 0$, i.e., $i \neq 1$, Eq. (4) is reduced to

$$\rho c_p \frac{T_{i,j+1} - T_{i,j}}{\Delta t} = \kappa \left( \frac{2}{i-1} \frac{T_{i+1,j} - T_{i,j}}{\Delta r^2} + \frac{T_{i+1,j} - 2T_{i,j} + T_{i-1,j}}{\Delta r^2} \right) + \rho_b c_{pb} \omega_b \left( T_c - T_{i,j} \right) + P_{i,j} \tag{C1}$$

where $P_{i,j}$ denotes the energy dissipation at the nodal point $i \cdot \Delta r$ at the time $j \cdot \Delta t$. Thus, $T_{i,j+1}$ can be computed from

$$T_{i,j+1} = T_{i,j} + \frac{\kappa \Delta t}{\rho c_p \Delta r^2} \left[ \left( 1 - \frac{2}{i-1} \right) \left( T_{i+1,j} - T_{i,j} \right) - \left( T_{i,j} - T_{i-1,j} \right) \right] + \frac{\rho_b c_{pb} \omega_b \Delta t}{\rho c_p} \left( T_c - T_{i,j} \right) + \frac{\Delta t}{\rho c_p} P_{i,j} \tag{C2}$$

for $i = 1,2,\cdots,M$ and $j = 1,2,\cdots,N$.

For $r = 0$, i.e., $i = 1$, we used the following L'Hopital's rule [23] to avoid dividing by zero:

$$\lim_{r \to 0} \frac{1}{r} \frac{\partial T}{\partial r} = \lim_{r \to 0} \frac{\partial^2 T}{\partial r^2} \tag{C3}$$

Then, we obtain

$$\rho c_p \frac{T_{1,j+1} - T_{1,j}}{\Delta t} = 6\kappa \frac{T_{2,j} - T_{1,j}}{\Delta r^2} + \rho_b c_{pb} \omega_b \left( T_c - T_{1,j} \right) + P_{1,j} \tag{C4}$$

or

$$T_{1,j+1} = T_{1,j} + \frac{6\kappa \Delta t}{\rho c_p \Delta r^2} \left( T_{2,j} - T_{1,j} \right) + \frac{\rho_b c_{pb} \omega_b \Delta t}{\rho c_p} \left( T_c - T_{1,j} \right) + \frac{\Delta t}{\rho c_p} P_{1,j} \tag{C5}$$

for $j = 1,2,\cdots,N$. For numerical stability, the following condition should be satisfied [24]:

$$\frac{\kappa \Delta t}{\rho c_p \Delta r^2} \leq 0.5 \tag{C6}$$

As boundary conditions, $\kappa\, \partial T / \partial r$ was taken as zero at the center and outer boundary, i.e., $T_{2,j} = T_{1,j}$ and $T_{M,j} = T_{M-1,j}$ for $j = 1,2,\cdots,N$. As initial conditions, the temperature at $t = 0$ was assumed to be $T_c$, i.e., $T_{i,1} = T_c$ for $i = 1,2,\cdots,M$.

## References

1. Abe M, Hiraoka M, Takahashi M, Egawa S, Matsuda C, Onoyama Y, Morita K, Kakehi M, Sugahara T. Multi-institutional studies on hyperthermia using an 8-MHz radiofrequency capacitive heating device (Thermotron RF-8) in combination with radiation for cancer therapy. Cancer. 1986; 58: 1589–95.
2. Oura S, Tamaki T, Hirai I, Yoshimasu T, Ohta F, Nakamura R, Okamura Y. Radiofrequency ablation therapy in patients with breast cancers two centimeters or less in size. Breast Cancer. 2007; 14: 48–54.
3. Seip R, Ebbini ES. Noninvasive estimation of tissue temperature response to heating fields using diagnostic ultrasound. IEEE Trans Biomed Eng. 1995; 42: 828–39.
4. Gilchrist RK, Medal R, Shorey WD, Hanselman RC, Parrott JC, Taylor CB. Selective inductive heating of lymph nodes. Ann Surg. 1957; 146: 596–606.
5. Jordan A, Scholz R, Maier-Hauff K, Johannsen M, Wust P, Nodobny J, Schirra H, Schmidt H, Deger S, Loening S, Lanksch W, Felix R. Presentation of a new magnetic field therapy system for the treatment of human solid tumors with magnetic fluid hyperthermia. J Magn Magn Mat. 2001; 225: 118–26.
6. Murase K, Oonoki J, Takata H, Song R, Angraini A, Ausanai P, Matsushita T. Simulation and experimental studies on magnetic hyperthermia with use of superparamagnetic iron oxide nanoparticles. Radiol Phys Technol. 2011; 4: 194–202.
7. Rosensweig RE. Heating magnetic fluid with alternating magnetic field. J Magn Magn Mat. 2002; 252: 370–4.
8. Neuberger T, Schopf B, Hofmann H, Hofmann M, von Rechenberga B. Superparamagnetic nanoparticles for biomedical applications: possibilities and limitations of a new drugdelivery system. J Magn Magn Mat. 2005; 293: 483–96.
9. Ito A, Shinkai M, Honda H, Kobayashi T. Medical applications of functionalized magnetic nanoparticles. J Biosci Bioeng. 2005; 100: 1–11.
10. Balivada S, Rachakatla RS, Wang H, Samarakoon TN, Dani RK, Pyle M, Kroh FO, Walker B, Leaym X, Koper OB, Tamura M, Chikan V, Bossmann SH, Troyer DL. A/C magnetic hyperthermia of melanoma mediated by iron(0)/iron oxide core/shell magnetic nanoparticles: a mouse study. BMC Cancer. 2010; 10: 119–27.
11. Pennes HH. Analysis of tissue and arterial blood temperatures in the resting human forearm. J Appl Physiol. 1948; 1:93–122.
12. Gao B, Langer S, Corry PM. Application of the time-dependent Green's function and

Fourier transforms to the solution of the bioheat equation. Int J Hyperthermia. 1995; 11: 267–85.

13. Durkee JW, Antich PP, Lee CE. Exact solutions to the multiregion time-dependent bioheat equation. I: solution development. Phys Med Bio. 1990; 35: 847–67.
14. Vyas R, Rustgi ML. Green's function solution to the tissue bioheat equation. Med Phys. 1992; 19: 1319–24.
15. Andra W, d'Ambly CG, Hergt R, Hilger I, Kaiser WA. Temperature distribution as function of time around a small spherical heat source of local magnetic hyperthermia. J Magn Magn Mat. 1999; 194: 197–203.
16. Deng Z-S, Liu J. Analytical study on bioheat transfer problems with spatial or transient heating on skin surface or inside biological bodies. J Biomech Eng. 2002; 124: 638–49.
17. Bagaria HG, Johnson DT. Transient solution to the bioheat equation and optimization for magnetic fluid hyperthermia treatment. Int J Hyperthermia. 2005; 21: 57–75.
18. Giordano MA, Gutierrez G, Rinaldi C. Fundamental solutions to the bioheatequation and their application to magnetic fluid hyperthermia. Int J Hyperthermia. 2010; 26: 475–84.
19. Ozisik MN. 'Boundary Value Problems of Heat Conduction', International Textbook Company, Scranton, Pennsylvania, 1968, pp. 43–124.
20. Carslaw HS, Jaeger JC. 'Conduction of Heat in Solids', Oxford at the Clarendon Press, Oxford, 1959, pp. 455–465.
21. Maenosono S, Saita S. Theoretical assessment of FePt nanoparticles as heating elements for magnetic hyperthermia. IEEE Trans Magn. 2006; 42: 1638–42.
22. Jordan A, Scholz R, Wust P, Fahling H, Krause J, Wlodarczyk W, Sander B, Vogl T, Felix R. Effects of magnetic fluid hyperthermia (MFH) on C3H mammary carcinoma in vivo. Int J Hyperthermia. 1997; 13: 587–605.
23. Spivak M. 'Calculus', Publish or Perish, Houston, 1994, pp. 201–202.
24. Smith GD. 'Numerical Solution of Partial Differential Equations with Exercises and Worked Solutions', Oxford University Press, London, 1965, p. 93.